\documentclass{article}
\usepackage{sprocl,graphicx,amsmath}

\def\ie{i.e.\ }
\def\eg{e.g.\ }
\def\alphas{\alpha_{\mathrm{s}}}
\def\mtpole{m_{\mathrm{t,pole}}}
\def\O{\mathcal{O}}
\def\M{\mathcal{M}}
\def\d{\mathrm{d}}
\def\eett{e^+e^-\to \bar tt}
\def\MBorn{\M_{\text{Born}}}
\def\Mloop{\M_{\text{1-loop}}}
\def\Mself{\M_{\text{self}}}

\def\Mverte{\M_{\text{vert},e}}
\def\Mvertt{\M_{\text{vert},t}}
\def\Mbox{\M_{\text{box}}}
\def\rest{{\text{rest}}}
\def\Re{\mathop{\mathrm{Re}}}

\begin{document}

\hfill hep-ph/9912262

\hfill TTP99-50

\hfill KA-TP-26-1999

\hfill BNL-HET-99/42

\bigskip

\title{Top production above threshold: \\
Electroweak and QCD corrections combined\footnote{Talk presented by J.\
K\"uhn at the Linear Collider Workshop, Sitges, Spain, April 1999.}}

\author{J.~K\"uhn$^1$, T.~Hahn$^2$, and R.~Harlander$^3$}

\address{$^1$Institut f\"ur Theoretische Teilchenphysik\\
$^2$Institut f\"ur Theoretische Physik \\
Universit\"at Karlsruhe,
D--76128 Karlsruhe, Germany \\[.5ex]
$^3$HET, Physics Department\\
Brookhaven National Laboratory, Upton, NY 11973}

\maketitle

\abstracts{Top quark production in electron--positron annihilation is one
of the benchmark reactions at a future linear collider. Both electroweak
and QCD corrections are large, amounting to 10\% or even more in specific
kinematic regions. In this note we present a method which allows to
combine the dominant terms from both sources, thus improving considerably
the result based on a simple addition of both corrections.}

\section{Introduction}

For a precise prediction of the continuum cross-section away from
threshold both the one-loop electroweak corrections~\cite{BeMH91} and the
QCD corrections of order $\alphas^2$~\cite{ChKS96} are needed. Both
contributions are sizeable, amounting to 10\% for the electroweak
corrections at 2 TeV and more than 100\% for the QCD corrections at 400
GeV.

While both corrections are available in the literature, so far no attempt
has been made to combine both. Simply adding the electroweak and QCD
corrections neglects the mixed $\O(\alpha\alphas)$ contributions, and
performing the complete $\O(\alpha\alphas)$ calculation is of course a
formidable task. So the question arises: Can one combine the available
pieces such that the dominant mixed $\O(\alpha\alphas)$ terms are taken
into account?

Unfortunately, not even this is straightforward because the QCD
corrections are as yet available only for the vector and axial-vector part
of the {\em correlator}, \ie they are given as a correction factor
multiplying the vector and axial-vector part of the {\it squared}
amplitude. It is not clear how to combine these results with those parts
of the electroweak amplitude which cannot simply be decomposed into a
vector and an axial-vector part. In this paper we discuss a strategy which
allows to combine these two ingredients in an approximate way without
evaluating the complete mixed corrections of order $\alpha\alphas$.

Purely electromagnetic corrections (apart from the initial-state radiation
which can be treated separately) are small and are hence omitted from this
preliminary analysis. They will be included in a future paper. Very close
to threshold the Coulomb enhancement (in other words, bound-state effects)
have to be taken into account. To avoid this region we adopt $2\mtpole +
10$ GeV as lower limit. A discussion of the matching between the
fixed-order treatment and the one based on Green functions including
Coulomb resummation can be found \eg in~\cite{HoT99}.

In the following sections we first discuss the electroweak and strong
effects separately and then present our recipe for combining both.

\section{Electroweak corrections}

\begin{figure}
\includegraphics[width=\linewidth]{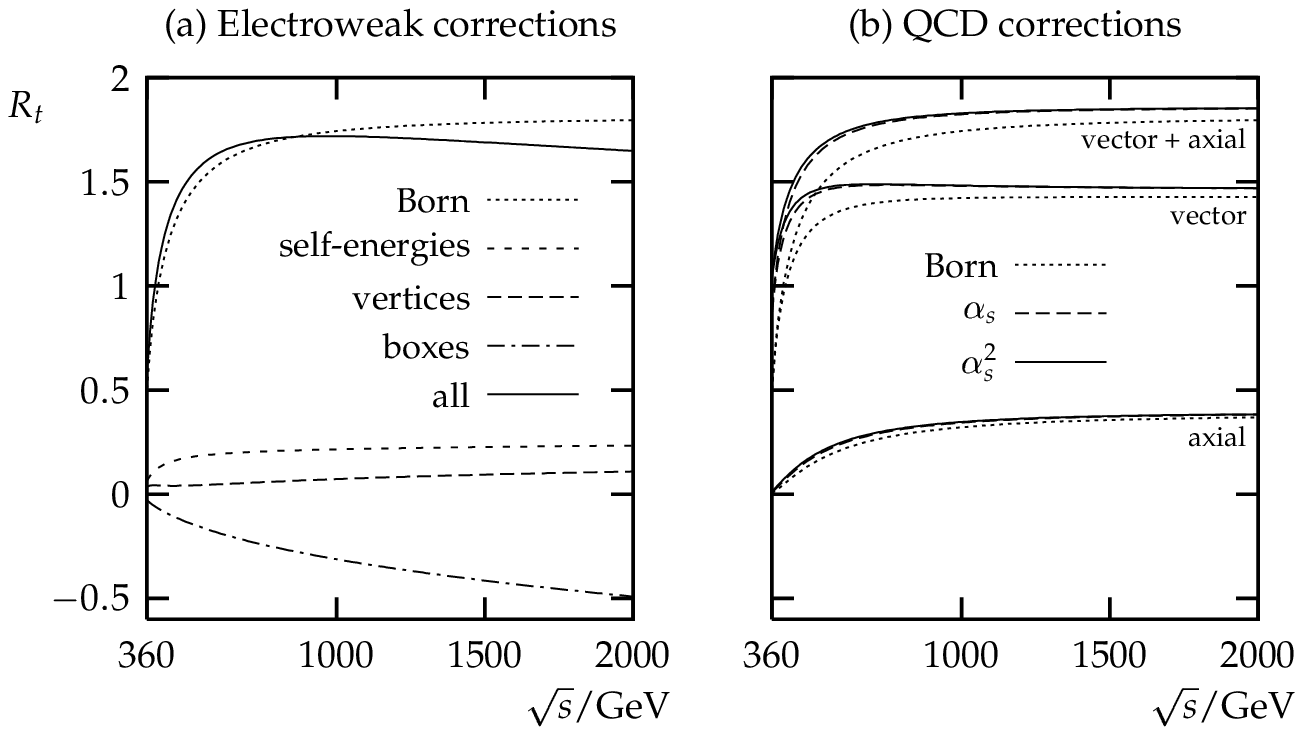}
\caption{\label{fig:ewqcdalone}The electroweak (a) and QCD (b)
corrections to $\eett$.}
\end{figure}

Figure \ref{fig:ewqcdalone} shows the total $\eett$ cross-section,
normalized to $\sigma_{\text{pt}} = 4\pi\alpha^2/3s$, where $s$ is the cms
energy squared.

The electroweak corrections are plotted in Fig.\ \ref{fig:ewqcdalone}(a),
where the dotted line is the Born approximation and the solid line is the
full $\O(\alpha)$ result. The one-loop diagrams can be separated into
self-energy insertions on the $\gamma$- and $Z$-propagator, corrections of
the electron and top vertex, and box contributions (see Fig.\
\ref{fig:ewqcdalone}(a)),
\begin{equation}
\Mloop = \Mself + \Mverte + \Mvertt + \Mbox\,.
\end{equation}
The box contributions are sizeable and largely compensate the self-energy
and vertex contributions.

These results were obtained using {\sl FeynArts} and
{\sl FormCalc}~\cite{FAFC}. They agree with the ones in
Ref.~\cite{BeMH91}.

\section{QCD corrections}

The QCD corrections are shown in Figure \ref{fig:ewqcdalone}(b). The
dotted, dashed, and solid lines denote the Born, $\O(\alphas)$, and
$\O(\alphas^2)$ approximation, respectively. Large corrections are
observed close to threshold, a remnant of the $1/\beta$ singularity of
the QCD corrections. These are particularly important for the vector
current and lead to a non-vanishing cross-section for small $\beta$.

Technically, the corrections are given by two factors $R_V$ and $R_A$ that
multiply the vector and axial-vector part of the lowest-order
cross-section, respectively.

\section{Combined electroweak and QCD corrections}

We now derive a strategy for the proper treatment of the terms of
order $\alpha (4\pi \alphas/\beta)$ which dominate in the region of small 
$\beta$, still without resummation of these terms, thus requiring formally
$4\pi\alphas/\beta\ll 1$. We start by decomposing the electroweak one-loop
amplitude into vector and axial-vector parts, for which we know how to
compute the QCD corrections, and the rest:
\begin{equation}
\begin{aligned}
\Mloop &= \Mloop^V + \Mloop^A + \Mloop^\rest \\[.5ex]
\llap{where\quad}
\Mloop^V &= \Mself^V + \Mverte^V + \Mvertt^V\,, \\
\Mloop^A &= \Mself^A + \Mverte^A + \Mvertt^A\,, \\
\Mloop^\rest &= \Mvertt^\rest + \Mbox\,.
\end{aligned}
\end{equation}
All terms with a $V$ superscript are proportional to $\bar u_t\gamma_\mu
v_t$ and all terms with an $A$ superscript are proportional to $\bar
u_t\gamma_5\gamma_\mu v_t$. The only corrections which have other Dirac
structures are the top vertex and the boxes. The QCD corrections to
$\Mloop^V$ and $\Mloop^A$ are identical to those to $\MBorn^V$ and
$\MBorn^A$, respectively, because they both have the same Dirac
structure.

Next, observe that terms of the form $\M^V\M^{A*}$ or $\M^A\M^{V*}$ drop
out of the squared amplitude after spin summation and angular integration.
This means that the cross-section can be written as
\begin{equation}
\begin{aligned}
\sigma \propto \sum_{\text{spins}} \int\d\Omega\,
\Bigl[
& \left|\MBorn^V\right|^2 + 
  2\Re \left(\Mloop^V + \Mloop^\rest\right)\MBorn^{V*} + \\[-1.5ex]
& \left|\MBorn^A\right|^2 +
  2\Re \left(\Mloop^A + \Mloop^\rest\right)\MBorn^{A*}
\Bigr].
\end{aligned}
\end{equation}
This decomposition is exact at ${\cal O}(\alpha)$. Now we tack on the
QCD corrections as
\begin{equation}
\label{eq:comb}
\begin{aligned}
\sigma \propto \sum_{\text{spins}} \int\d\Omega\,
\Bigl[
& R_V \left(
  \left|\MBorn^V\right|^2 + 
  2\Re \left(\Mloop^V + \Mloop^\rest\right)\MBorn^{V*}\right) + \\[-1.5ex]
& R_A \left(
  \left|\MBorn^A\right|^2 +
  2\Re \left(\Mloop^A + \Mloop^\rest\right)\MBorn^{A*}\right)
\Bigr].
\end{aligned}
\end{equation}
Of course, this equation does not really reproduce the contributions of
order $\alpha\alphas$ and $\alpha\alphas^2$. Close to threshold, however,
the s-wave contribution is entirely contained in the terms that are
proportional to $\MBorn^V$. They dominate over the $\MBorn^A$-terms by a
factor of $1/\beta^2$. This means that for not-too-large $\beta$, the
electroweak corrections act at distances that are short compared to the
inverse relative $t$--$\bar t$ momentum, and so the QCD corrections, in
particular the Coulomb enhancement, can be included as an overall factor
multiplying the vector part $\MBorn^{V*}$.

\begin{figure}
\centerline{\includegraphics[width=.75\linewidth]{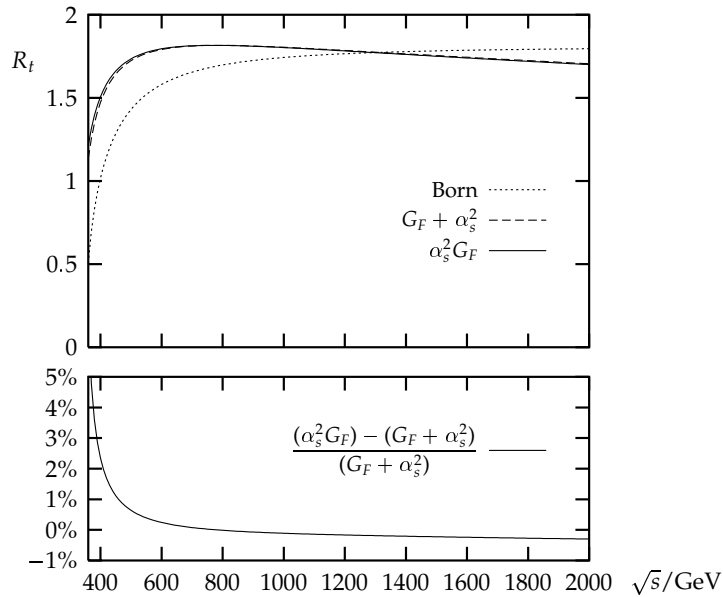}}
\caption{\label{fig:ewqcd}{\it Upper part:} Comparison of the naive
addition of electroweak and QCD corrections (solid line) with the
approximate way of including the dominant $\O(\alpha\alphas)$
contributions in Eq.\ \eqref{eq:comb} (dashed line).
{\it Lower part:} The relative difference of the two methods.}
\end{figure}

The results of Eq.\ \eqref{eq:comb} are plotted in Fig.\ \ref{fig:ewqcd}
(solid line). To indicate the significance of this prescription we also
give the prediction obtained by simply adding QCD and electroweak
corrections (dashed line). As expected, the difference is largest in the
low-energy region, amounting to 5\% at 360 GeV.

\section*{Acknowledgments}

We thank C.~Schappacher for cross-checking the electroweak results and
M.~Steinhauser for providing us with the Pad\'e results~\cite{ChKS96} of
the $\alpha_s^2$ corrections to $R_V$ and $R_A$.


\begin{thebibliography}{9}

\bibitem{BeMH91}
W.~Beenakker, S.C.~van der Marck, and W.~Hollik,
  {\sl Nucl.\ Phys.} {\bf B365} (1991) 24.

\bibitem{ChKS96}
K.G.~Chetyrkin, J.H.~K\"uhn, and M.~Steinhauser,
  {\sl Nucl.\ Phys.} {\bf B482} (1996) 213.
K.G.~Chetyrkin, R.~Harlander, J.H.~K\"uhn, and M.~Steinhauser,
  {\sl Nucl.\ Phys.} {\bf B503} (1997) 339.
K.G.~Chetyrkin, J.H.~K\"uhn, and M.~Steinhauser,
  {\sl Nucl.\ Phys.} {\bf B505} (1997) 40.
R.~Harlander and M.~Steinhauser, {\sl Eur.\ Phys.\ J.} {\bf C2} (1998) 151.

\bibitem{HoT99}
A.H.~Hoang and T.~Teubner, {\sl Phys.\ Rev.} {\bf D60} (1999) 114027.

\bibitem{FAFC}
J.~K\"ublbeck, M.~B\"ohm, and A.~Denner,
  {\sl Comp.\ Phys.\ Commun.} {\bf 60} (1990) 165.
T.~Hahn and M.~P\'erez-Victoria,
  {\sl Comp.\ Phys.\ Commun.} {\bf 118} (1999) 153.

\end{thebibliography}
\end{document}